\pdfoutput=1

\documentclass[aps,prb,reprint,twocolumn]{revtex4-2}

\usepackage{amsfonts,amsmath,amssymb}
\usepackage{bbold}
\usepackage{bm}
\usepackage{mathrsfs}
\usepackage{color}
\usepackage{xcolor}
\usepackage{graphicx}
\usepackage{placeins}
\usepackage{mhchem}
\usepackage{nicefrac}
\usepackage{mathtools}
\usepackage{booktabs}
\usepackage{siunitx}
\usepackage{url}
\usepackage{fourier}
\usepackage[most]{tcolorbox}
\usepackage{lipsum}

\newcommand{\opcn}[5]{\hat{#1}_{\bm{#3}#4,#5}^{#2 \dagger}(t)}
\newcommand{\opan}[5]{\hat{#1}_{\bm{#3}#4,#5}^{#2 \vphantom{\dagger}}(t)}
\newcommand{\funcn}[5]{#1_{\bm{#3}#4,#5}^{#2 \vphantom{\dagger}}}
\newcommand{\funcastn}[5]{#1_{\bm{#3}#4,#5}^{#2 \ast \vphantom{\dagger}}}
\newcommand{\eps}[3]{\varepsilon_{\bm{#1},#2}^{#3 \vphantom{\dagger}}}

\newcommand{\bc}[1]{\hat{b}_{#1 \vphantom{\bm{k}}}^{\dagger}}
\newcommand{\ba}[1]{\hat{b}_{#1 \vphantom{\bm{k}}}^{\vphantom{\dagger}}}
\newcommand{\bcvec}[1]{\hat{b}_{\bm{#1}}^{\dagger}}
\newcommand{\bavec}[1]{\hat{b}_{\bm{#1}}^{\vphantom{\dagger}}}

\newcommand{\func}[4]{#1_{\bm{#3},#4}^{#2 \vphantom{\dagger}}}
\newcommand{\funcast}[4]{#1_{\bm{#3},#4}^{#2 \ast \vphantom{\dagger}}}

\begin{document}


\title{Material-realistic modelling of quantum many-body effects in a monolayer TMDC nanolaser device}


\author{J. Buchgeister}
\email[email:]{jbuchgeister@itp.uni-bremen.de}
\affiliation{Institut für Theoretische Physik, Universität Bremen, Otto-Hahn-Allee 1, 28359 Bremen}
\affiliation{Bremen Center for Computational Materials Science BCCMS, University of Bremen, Am Fallturm 1, 28359 Bremen}
\author{A. Steinhoff}
\affiliation{Institut für Theoretische Physik, Universität Bremen, Otto-Hahn-Allee 1, 28359 Bremen}
\affiliation{Bremen Center for Computational Materials Science BCCMS, University of Bremen, Am Fallturm 1, 28359 Bremen}
\author{D. Erben}
\affiliation{Institut für Theoretische Physik, Universität Bremen, Otto-Hahn-Allee 1, 28359 Bremen}
\affiliation{Bremen Center for Computational Materials Science BCCMS, University of Bremen, Am Fallturm 1, 28359 Bremen}
\author{M. Lorke}
\affiliation{Institut für Theoretische Physik, Universität Bremen, Otto-Hahn-Allee 1, 28359 Bremen}
\affiliation{Bremen Center for Computational Materials Science BCCMS, University of Bremen, Am Fallturm 1, 28359 Bremen}
\author{F. Jahnke}
\affiliation{Institut für Theoretische Physik, Universität Bremen, Otto-Hahn-Allee 1, 28359 Bremen}
\affiliation{Bremen Center for Computational Materials Science BCCMS, University of Bremen, Am Fallturm 1, 28359 Bremen}


\date{September 26th 2024}

\begin{abstract}
The efficient light-matter interaction in combination with the small volume occupied by monolayer transition-metal dichalcogenides (TMDCs) makes this material class a notable option as gain layer
in future opto-electronic devices. Many-body effects of excited carriers influence the emission dynamics due to the introduction of optical non-linearities following excitation, but the exact
mechanisms remain unexamined from a theoretical point of view. In this paper, we present a material-realistic microscopic theory of a device based on an MoS$_2$-monolayer, which demonstrates
stimulated emission activity at room temperature. The modelling procedure combines Coulomb and light-matter interaction matrix elements with doublet-level Quantum
Laser Equations (QLEs). These give access to the dynamics of the photon-assisted polarisation, populations, and photon number while allowing the solution of a multi-scale problem ranging from
femto- to nanoseconds. The input-output curve, hole burning and spectral clamping obtained from this theory present strong indications of electron-hole-plasma-based lasing occurring at densities
above $5 \cdot 10^{13}\,\text{cm}^{-2}$ in this device.
\end{abstract}


\maketitle


\section{Introduction}

Global power consumption associated with data centres is greatly increasing, with the European Union alone being projected to reach levels in 2026 which are $30\,\%$ above those in 2023,
most of which can be attributed to increased expenditures in digitalisation and artifical intelligence \cite{avelar2023ai,cam2024electricity}.
The incorporation of more energy efficient optoelectronic devices and on-chip photonics for the short-distance optical links in the data industry can help counteract this development.
One step in the realisation of such devices is the development of nanoscale sources of stimulated emission, in particular those operating at room temperature.
Two-dimensional transition-metal dichalcogenides (TMDCs) present an interesting material class for such nanolaser devices due to the small volume they occupy as well as the possibility to
fabricate monolayers of gain material on top of photonic crystal cavities allowing both, scalability as well as compatibility with integrated photonics \cite{wu2015monolayer}.

\begin{figure}[h!]
\vspace{0.00cm}
\includegraphics[width=0.40\textwidth]{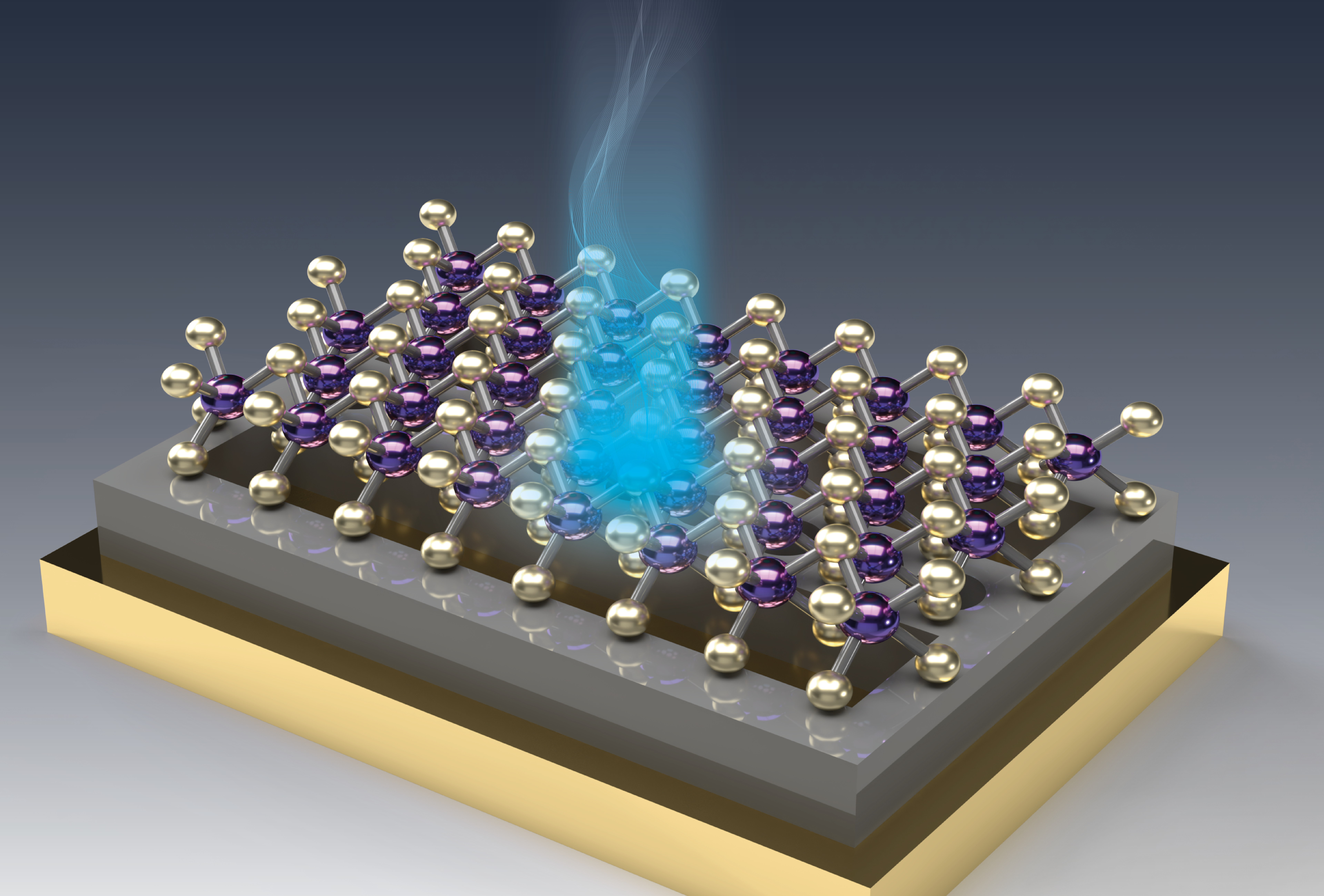}
\caption{Visualisation of the TMDC nanolaser: a monolayer of MoS$_2$ on an SiO$_2$ photonic crystal cavity.}
\label{fig:A}
\vspace{0.00cm}
\end{figure}

While previous works \cite{wu2015monolayer,ye2015monolayer,li2017room,liu2022room} have attributed an excitonic origin to the TMDC stimulated emission, in this paper we will explore how
a fully quantum mechanical theory considering many-body effects and the efficient light-matter interaction \cite{salehzadeh2015optically,schneider2018two} results in electron-hole-plasma-based
lasing for a monolayer of MoS$_2$ as gain material on an SiO$_2$ photonic crystal cavity.
Microscopically modelling the dynamics inherent to a nanolaser utilising a monolayer TMDC as active gain material requires a multi-level approach adressing a variety of effects
with sufficient detail: Proceeding from a material-realistic lattice Hamilton operator in two dimensions, dipole and Coulomb matrix elements can be calculated.
The Coulomb matrix elements have to consider both, dielectric screening due to the material and substrate layer as well as excitation-induced screening as a consequence of populations
being driven by an optical pump pulse. The excited carrier contributions result in band structure renormalisations and screening which in turn greatly influence the temporal
evolution of the governing Quantum Laser Equations (QLEs). Carrier population dynamics are furthermore influenced by relaxation effects due to carrier-carrier and carrier-phonon
scattering as well as spontaneous recombination into emission channels which, as opposed to the lasing mode, have no cavity feedback. Quantum-optical modelling of gain layer and cavity also
requires the consideration of resonantor losses as well as dephasing effects.
Solving these dynamics whilst maintaining a high level of microscopic detail is particularly challenging due to the inherent multi-scale nature of the problem.
While Coulomb-driven dynamics and energy band renormalisations are to be resolved on a femtosecond timescale, the lasing dynamics operate on the nanosecond timescale.
Here, we will present a method that handles this property of the problem by introducing effective modelling when suitable in order to realise reasonable computation times
without sacrificing microscopic precision.

\section{Theory}

The Quantum Laser Equations are derived from the Heisenberg Equation of Motion (EoM) approach \cite{carmichael1993open,carmichael1999statistical} for the expectation values of observables
which are composed of photon and carrier operators. Here, \smash{$\hat{b}^{\dagger}_{\bm{\xi}}$} is a photon creation operator for mode $\bm{\xi}$ and \smash{$\opcn{a}{\lambda}{k}{}{s}$}
an electron creation operator for energy band $\lambda$, momentum $\bm{k}$, and spin $s$. Lindblad terms \cite{lindblad1976generators,breuer2002theory} are calculated in conjunction with this
procedure in order to model resonator losses by means of the Lindblad generator
\smash{$\hat{V}^{\dagger} = (2 \hbar \kappa^{\vphantom{\dagger}}_{\bm{\xi}})^{-\nicefrac{1}{2}} \hat{b}^{\dagger}_{\bm{\xi}}$} where \smash{$\kappa^{\vphantom{\dagger}}_{\bm{\xi}}$} is the
associated resonator loss rate in \smash{ps$^{-1}$}. For a quantum mechanical operator $\hat{A}$, e.g. the photon number, this reads:

\begin{align}
	{\frac{\text{d}}{\text{d}t}} \langle \hat{A} \rangle
	&=
	{\frac{i}{\hbar}}
	\langle \left[
	\hat{\mathcal{H}}_{\text{car.}} + \hat{\mathcal{H}}_{\text{pho.}}
	+ \hat{\mathcal{H}}_{\text{int.}} + \hat{\mathcal{H}}_{\text{Cou.}}, \hat{A}
	\right] \rangle
\nonumber \\
	&+ {\frac{i}{\hbar}} \langle \hat{V}^{\dagger} \hat{A} \hat{V} \rangle
	- {\frac{i}{2 \hbar}} \langle \left\{ \hat{V}^{\dagger} \hat{V}, \hat{A} \right\} \rangle
\quad.
\end{align}

While previous works \cite{chow2014emission,koulas2022quantum} have employed a similar technique using the contributions of the carriers, photons, and the light-matter interaction
with a fully quantised light field, a unique aspect of the research presented here, which is not yet established in the field of quantum mechanical nanolaser modelling, is the
additional consideration of inherent many-body effects such as the Coulomb interactions between excited carriers.
The associated matrix elements are calculated based on a lattice Hamilton operator \smash{$\hat{\mathcal{H}}_{\text{TB} | \bm{k},s}^{\alpha,\beta}$} stated in a
basis consisting of Wannier orbital functions \smash{$| w_{\bm{R}_{j},s}^{\alpha \vphantom{\dagger}} \rangle$} localised around the transition metal's dominant $d$-orbtials
$\alpha$ at lattice positions \smash{$\bm{R}_{j}$} \cite{liu2013three}.
The lattice Hamilton operator is obtained from a tight binding model that considers third-nearest-neighbour hoppings as well as on-site spin-orbit coupling (SOC) with parameters fitted
to first-principles energy bands in the generalised-gradient approximation (GGA).
Expanding the Bloch states \smash{$| \func{\phi}{\lambda}{k}{s} \rangle$} of the associated tight binding model, which are distinct in band index $\lambda$, momentum $\bm{k}$,
and spin $s$, into Wannier orbitals leads to the occurrence of the expansion coefficients \smash{$\func{c}{\alpha,\lambda}{k}{s}$}:

\begin{align}
	| \func{\phi}{\lambda}{k}{s} \rangle
	= \sum_{\alpha} \func{c}{\alpha,\lambda}{k}{s} \frac{1}{\sqrt{N}} \sum_{j = 1}^{N} e^{i \bm{k} \cdot \bm{R}_{j}} | w_{\bm{R}_{j},s}^{\alpha \vphantom{\dagger}} \rangle
\quad.
\end{align}

The general form of the Coulomb Hamilton operator in the Bloch state representation with the associated coefficients then reads in second quantisation
(compare \cite{steinhoff2021microscopic}):

\begin{align}
	\hat{\mathcal{H}}_{\text{Cou.}}
	&=
	\frac{1}{2} \sum_{s,s' \vphantom{\bm{k}}} \sum_{\bm{k},\bm{k'},\bm{q}} \sum_{\lambda_{1},\lambda_{2},\lambda_{3},\lambda_{4} \vphantom{\bm{k}}}
	\opcn{a}{\lambda_{1}}{k}{}{s} \opcn{a}{\lambda_{2}}{k'}{}{s'} \opan{a}{\lambda_{3}}{k'}{+\bm{q}}{s'} \opan{a}{\lambda_{4}}{k}{-\bm{q}}{s}
\nonumber \\
	&\times
	\mathcal{V}_{\bm{k},\bm{k'},\bm{k'}+\bm{q},\bm{k}-\bm{q},s,s'}^{\lambda_{1},\lambda_{2},\lambda_{3},\lambda_{4} \vphantom{\dagger}}
\quad,
\end{align}

with:

\begin{align}
	\mathcal{V}_{\bm{k},\bm{k'},\bm{k'}+\bm{q},\bm{k}-\bm{q},s,s'}^{\lambda_{1},\lambda_{2},\lambda_{3},\lambda_{4} \vphantom{\dagger}}
	= \sum_{\alpha,\beta} \funcastn{c}{\alpha,\lambda_{1}}{k}{}{s} \funcastn{c}{\beta,\lambda_{2}}{k'}{}{s'}
	\funcn{c}{\beta,\lambda_{3}}{k'}{+\bm{q}}{s'} \funcn{c}{\alpha,\lambda_{4}}{k}{-\bm{q}}{s}
	\mathcal{V}_{\bm{q},s,s'}^{\alpha,\beta,\beta,\alpha \vphantom{\dagger}}
\quad.
\end{align}

This form considers momentum conservation with umklapp processes, hence all present momenta are originating from the first Brillouin zone with $\bm{q}$ characterising the momentum transfer
associated with the interaction. The employed orbital basis has been chosen to best represent the particular subspace which includes the optically relevant highest valence and lowest conduction
band denoted by $\lambda$. The focus on these energy bands is due to the fact that the Coulomb interaction is embedded in a nanolaser theory. The number of Coloumb matrix elements is furthermore
reduced by including only density-density like matrix elements associated with processes conserving the number of intergap excitations, an approach that focuses on the dominant matrix elements for
carrier interaction effects.
The microscopic matrix elements \smash{$\mathcal{V}_{\bm{q},s,s'}^{\alpha,\beta,\beta,\alpha \vphantom{\dagger}}$} are assumed spin-independent and averaged over all elements in the
orbital representation, hence only the quasi-momentum dependence remains, leading to \smash{$\mathcal{V}_{\bm{q},s,s'}^{\alpha,\beta,\beta,\alpha \vphantom{\dagger}}
= \mathcal{V}_{\bm{q}}^{\vphantom{\dagger}}$}. This is equivalent to neglecting local field effects and justified for an electron-hole plasma consisting of quasi-free carriers \cite{hanke1975local}.
However, the dominant orbital dependence remains by maintaining the expansion coefficients.
The particularly strong Coulomb interaction between excited carriers in TMDC monolayers originates from the layer's two-dimensionality: firstly due to the carrier confinement and secondly due to
the weak dielectric screening that has been the subject of various detailed studies \cite{steinhoff2014influence,erben2018excitation,steinhoff2018biexciton,steinke2020coulomb,erben2022optical}.
The model presented here caters to two different kinds of screening: background dielectric screening of the unexcited semiconductor and photonic crystal as well as the rather strong screening
from excited carriers following an optical excitation of the gain material. Since those excited carriers contributing to the emission of light accumulate in the valleys of the optically relevant
valence and conduction band, band structure renormalisations have to be considered for non-zero densities. A well-established method for modelling this effect is the screened-exchange Coulomb-hole
(SX-CH) approximation \cite{steinhoff2015efficient,pogna2016photo,erben2018excitation}.
When employing this method, three different types of Coulomb matrix elements occur: an unscreened bare Coulomb potential \smash{$U_{\bm{q}}^{\vphantom{\dagger}}$}, a background-screened potential
\smash{$V_{\bm{q}}^{\vphantom{\dagger}}$}, and a fully screened potential \smash{$W_{\bm{q}}^{\vphantom{\dagger}}$}.
Calculating the latter one is highly cost-intensive from a numerical point of view due to the necessity of re-calculating multiple matrices with momentum resolution: the many-body
effects require a resolution on the femtosecond timescale, however laser dynamics only reach convergence on the nanosecond timescale, in particular below the lasing threshold.
The numerical effort can be significantly reduced by employing suitable simplifications which still correctly represent the relevant physical processes.
The first one of these simplifications is the consideration of background screening by using a Keldysh-type potential for quasi-2D nanostructures which accounts for screening from surrounding
media from above (vacuum) and below (SiO$_2$) by means of a screening length \smash{$\varrho_0 = 2 \pi \kappa^{-1} \chi_{\text{2D}}$} that depends on the 2D planar polarisability
\smash{$\chi_{\text{2D}}$} of the gain material \cite{keldysh2023coulomb,cudazzo2011dielectric,berghauser2014analytical,kylanpaa2015binding} and the average dielectric constant of
the environment $\kappa$:

\begin{align}
	V_{\bm{q}}^{\vphantom{-1}}
	= \frac{1}{\mathcal{A}} \frac{e^2}{\varepsilon_{0} \varepsilon_{s}} \frac{1}{|\bm{q}|} (1 + \varrho_{0} |\bm{q}|)^{-1}
\quad.
\end{align}

Here, $\mathcal{A}$ is the normalisation area associated with the gain layer area. The 2D dielectric function exhibits a momentum dependence which allows for a greater potential with
less screening to occur at long wavelengths. For shorter wavelengths, however, the resulting potential is reduced, implying that the screening effect due to the material's polarisation
becomes dominant. The dielectric function of the excited carrier screening is calculated in the Debye limit. This static long-wavelength-limit subsequently results in the screening
contributions of excited carriers being dependent on their respective effective masses (assuming valley isotropy) and populations \cite{erben2018excitation}.
Here, $v$ and $c$ are the indices of the two optically relevant energy bands and $\nu_{v/c} = \{ \bm{k}_{\nu_{v/c}},s \}$ indicates the respective valley.
\smash{$\tilde{V}_{\bm{q}} = \mathcal{A} V_{\bm{q}}$} is a rescaled Keldysh-type potential:

\begin{align}
	W_{\bm{q}}^{\vphantom{-1}}
	= V_{\bm{q}}^{\vphantom{-1}} \left( 1 + \tilde{V}_{\bm{q}}^{\vphantom{-1}}
	\frac{\sum_{\nu_{v}} m_{\nu_{v}}(1-f_{\nu_{v}}) + \sum_{\nu_{c}} m_{\nu_{c}}f_{\nu_{c}}}{2 \pi \hbar^2} \right)^{-1}
\quad.
\end{align}

Besides the influence of excited carrier dynamics, the interaction between light and matter also has to be described consistently with other microscopic properties, such as the
energy landscape across the momentum space. Applying the G{\"o}ppert-Mayer transformation to the general form of the light-matter interaction Hamilton operator creates a pure
dipole as well as a bare Coulomb-like dipole-dipole contribution \cite{cohen1989photons,cohen1992atom,kira1999quantum}.
The latter is neglected here since previous research \cite{qiu2015nonanalyticity} has shown that in TMDCs such electron-hole exchange interactions only contribute to the
fine-structure splitting between like-spin (bright) and unlike-spin (dark) states, which exhibit a slightly lower excitation energy, hence only influencing the dynamics by
resulting in carrier loss to optically dark states. In the context of modelling nanolaser dynamics, which only consider spin- and momentum-diagonal state transition contributions
for lasing operation, this negligence results in slightly less excitation to be required for obtaining the desired emission response from the gain material.
The general form of the light-matter interaction Hamilton operator, utilising the same Bloch state representation as introduced for the Coulomb Hamilton operator, then reads in
second quantisation:

\begin{align}
	\hat{\mathcal{H}}_{\text{l.m.}}
	= - \sum_{\bm{k},s} \sum_{\lambda,\lambda'} \opcn{a}{\lambda}{k}{}{s} \opan{a}{\lambda'}{k}{}{s}
	\langle \func{\phi}{\lambda}{k}{s}
	\left| e \bm{r} \cdot \varepsilon_0^{-1} \bm{D}(\bm{r},t) \right| \func{\phi}{\lambda'}{k}{s} \rangle
\quad.
\end{align}

Applying the formal definition of the electromagnetic field post G{\"o}ppert-Mayer transformation as stated in \cite{cohen1989photons,kira1999quantum}, as well as only considering interband
transitions ($\lambda \neq \lambda'$), while neglecting intraband contributions ($\lambda = \lambda'$) due to the focus on those interactions with optical contributions, allows to express the
light-matter interaction Hamilton operator in second quantisation in terms of the dipole of the material. Here, \smash{$\eps{k}{s}{\lambda}$} is the energy band structure value (without applied
energy renormalisations), \smash{$\varepsilon_{\bm{\xi}} = (0.5 \hbar \omega_{\bm{\xi}} \varepsilon_0^{-1})^{\nicefrac{1}{2}}$} is the vacuum field amplitude, \smash{$\omega_{\bm{\xi}}$} the mode
frequency, \smash{$\bm{u}_{\bm{\xi}}(z)$} the mode function at position $z$ (with the gain layer being located at $z = Z_{0}$), and \smash{$\tilde{\mathcal{A}}$} the mode-associated normalisation
area.

\begin{align}
	\hat{\mathcal{H}}_{\text{l.m.}}
	&=
	i \hbar \sum_{\bm{k},s} \sum_{\lambda \neq \lambda' \vphantom{\bm{k}}} \sum_{\bm{\xi} \vphantom{\bm{k}}}
	\opcn{a}{\lambda}{k}{}{s} \opan{a}{\lambda'}{k}{}{s}
\nonumber \\
	&\times \left[ \bcvec{\xi}(t) \frac{1}{\hbar} \varepsilon_{\bm{\xi}} \frac{\bm{u}_{\bm{\xi}}^{\ast}(Z_0)}{\sqrt{\tilde{\mathcal{A}}}}
	- \bavec{\xi}(t) \frac{1}{\hbar} \varepsilon_{\bm{\xi}} \frac{\bm{u}_{\bm{\xi}}(Z_0)}{\sqrt{\tilde{\mathcal{A}}}} \right]
	\cdot \func{\bm{d}}{\lambda,\lambda'}{k}{s}
\quad,
\end{align}

with:

\begin{align}
	\func{\bm{d}}{\lambda,\lambda'}{k}{s}
	&=
	\frac{e}{i} \frac{1}{\func{\varepsilon}{\lambda}{k}{s} - \func{\varepsilon}{\lambda'}{k}{s}}
	\sum_{\alpha,\beta} \funcast{c}{\alpha,\lambda\vphantom{\beta}}{k}{s} \func{c}{\beta,\lambda'}{k}{s} \bm{\nabla}_{\bm{k}}^{\vphantom{\dagger}}
	\hat{\mathcal{H}}_{\text{TB} | \bm{k},s}^{\alpha,\beta}
\quad.
\end{align}

Calculating the explicit mode functions requires solving Maxwell's equations for a specific design of the photonic crystal and for various modes.
The purpose of said functions in the context of light-matter interaction is to provide a scaling factor for this interaction while the momentum-dependence is mediated via the
microscopic dipole of the gain material. However, the spatial dependence of the mode functions is not relevant since only their value at the position of the TMDC layer, where the
interaction occurs, is required. Thus it is legitimate to reduce computational effort and only explicitly consider the polarisations (here circular polarisation $\bm{\sigma}_{\bm{\xi}}$
as an orthonormal basis of the emission is assumed) and introduce an effective scaling parameter \smash{$g_{0,\bm{\xi}}$} which results in the definition of the overall light-matter
interaction strength:

\begin{align}
	\frac{1}{\hbar} \varepsilon_{\bm{\xi}} \frac{\bm{u}_{\bm{\xi},2D}^{\ast}(Z_0)}{\sqrt{\tilde{\mathcal{A}}}}
	\cdot \tilde{\bm{d}}_{\bm{k},s}^{\lambda,\lambda'\vphantom{\dagger}}
	\approx
	g_{0,\bm{\xi}} \left( \bm{\sigma}_{\bm{\xi}} \cdot \tilde{\bm{d}}_{\bm{k},s}^{\lambda,\lambda'\vphantom{\dagger}} \right)
	\equiv
	g_{\bm{k},s,\bm{\xi}}^{\lambda,\lambda'\vphantom{\dagger}}
\quad.
\end{align}

The theory presented in this work considers the interaction of a single lasing mode \smash{$\bm{\xi} = \bm{\xi_{L}}$} with the TMDC monolayer gain material. Emission, however, will be a mixture of
left- and right-polarised light, hence the Quantum Laser Equations presented below maintain a mode resolution; the mode index \smash{$\bm{\xi_{L}}$} thus implies not the occurrence of different
lasing frequencies, but caters to the existence of the two polarisation directions which have to be considered for one lasing frequency.
For a model that only utilises the two optically relevant energy bands, the execution of the band-index sum in the formal definition of the light-matter interaction Hamilton operator
\smash{$\hat{\mathcal{H}}_{\text{l.m.}}$} and neglecting off-resonant terms then results in two remaining summands. \newline
The inherent multi-scale nature of the problem requires fast carrier dynamics, i.e. the Coulomb interaction as well as the energy band renormalisations, to be resolved on a femtosecond timescale,
but allows those processes directly contributing to the lasing dynamics, which move towards a local equilibrium on a picosecond (above the lasing threshold) or nanosecond (below the lasing
threshold) timescale to be modelled effectively. Consistent modelling of resonator losses has been realised by means of Lindblad terms as introduced above. Similar to previous
works \cite{chow2014emission,koulas2022quantum}, laser dynamics such as the spontaneous recombination into those emission channels without cavity feedback has been incorporated by means of an
effective radiative loss rate \smash{$\gamma_{nl}$}; pumping of the gain material has been modelled with an optical pump of Gaussian shape \smash{$F_{\bm{k},s}^{p}$} centered at $2.5\,\text{eV}$
and effective pump rate $P$;
scattering due to carrier-carrier and carrier-phonon interactions enters the final set of equations via a single dephasing rate $\Gamma$. Furthermore, the relaxation-time approximation,
which allows excited carrier populations to evolve towards equilibrated Fermi distributions \smash{$F_{\bm{k},s}^{D,\lambda}$} at rate \smash{$\gamma_{rel}$}, is introduced.
This procedure eventually yields the Quantum Laser Equations on the doublet level, a set of coupled differential equations for the photon-assisted polarisation
\smash{$\psi^{}_{\bm{\xi_{L}},\bm{k},s}(t)$}, the populations of the optically relevant energy bands, \smash{$f^{c}_{\bm{k},s}(t)$} and \smash{$f^{v}_{\bm{k},s}(t)$}, as well as the overall photon
number \smash{$n(t)$}, which are formally defined as:

\vspace{-0.25cm}
\begin{align}
	\psi^{}_{\bm{\xi_{L}},\bm{k},s}(t) &= \langle \bc{\bm{\xi_{L}}}(t) \opcn{v}{}{k}{}{s} \opan{c}{}{k}{}{s} \rangle
\quad, \\
	f^{c}_{\bm{k},s}(t) &= \langle \opcn{c}{}{k}{}{s} \opan{c}{}{k}{}{s} \rangle
\quad, \\
	f^{v}_{\bm{k},s}(t) &= \langle \opcn{v}{}{k}{}{s} \opan{v}{}{k}{}{s} \rangle
\quad, \\
	n(t) &= \sum_{\bm{\xi_{L}}} \langle \bc{\bm{\xi_{L}}}(t) \ba{\bm{\xi_{L}}}(t) \rangle
\quad.
\end{align}

\begin{widetext}
\vspace{-0.50cm}
\begin{align}
	\frac{\text{d}}{\text{d}t} \psi^{}_{\bm{\xi_{L}},\bm{k},s}(t)
		=
		&\phantom{+} \frac{i}{\hbar}
		\left( \left[
		\eps{k}{s}{v}
		+ \sum_{\bm{k'}} (1 - f^{v}_{\bm{k'},s}(t))
		W_{\bm{k'},\bm{k},\bm{k'},\bm{k},s,s}^{v,v,v,v \vphantom{\dagger}}
		\right.
		- \frac{1}{2} \sum_{\bm{k'}}
		\left[ W_{\bm{k'},\bm{k},\bm{k'},\bm{k},s,s}^{v,v,v,v \vphantom{\dagger}} - V_{\bm{k'},\bm{k},\bm{k'},\bm{k},s,s}^{v,v,v,v \vphantom{\dagger}} \right]
		- \sum_{\bm{k'}} f^{c}_{\bm{k'},s}(t)
		U_{\bm{k'},\bm{k},\bm{k'},\bm{k},s,s}^{c,v,c,v \vphantom{\dagger}}
		\right]
	\nonumber \\
		\phantom{=}
		&{-} \left[
		\eps{k}{s}{c}
		- \sum_{\bm{k'}} f^{c}_{\bm{k'},s}(t)
		W_{\bm{k'},\bm{k},\bm{k'},\bm{k},s,s}^{c,c,c,c \vphantom{\dagger}}
		+ \left. \frac{1}{2} \sum_{\bm{k'}}
		\left[ W_{\bm{k'},\bm{k},\bm{k'},\bm{k},s,s}^{c,c,c,c \vphantom{\dagger}} - V_{\bm{k'},\bm{k},\bm{k'},\bm{k},s,s}^{c,c,c,c \vphantom{\dagger}} \right]
		+ \sum_{\bm{k'}} (1 -f^{v}_{\bm{k'},s}(t))
		U_{\bm{k'},\bm{k},\bm{k'},\bm{k},s,s}^{v,c,v,c \vphantom{\dagger}}
		\right] + \hbar \omega_{\bm{\xi_{L}}} \right)
		\psi^{}_{\bm{\xi_{L}},\bm{k},s}(t)
	\nonumber \\
		\phantom{=}
		&{+} \left( g_{\bm{k},s,\bm{\xi_{L}}}^{v,c \vphantom{\dagger}} \right)^{\ast} \left( f^{c}_{\bm{k},s}(t)
		\left( 1 - f^{v}_{\bm{k},s}(t) \right) \right)
		+ \left( f^{v}_{\bm{k},s}(t) - f^{c}_{\bm{k},s}(t) \right)
		\left(
		\frac{i}{\hbar} \sum_{\bm{k'}}
		\left(\psi^{}_{\bm{\xi_{L}},\bm{k'},s}(t)
		W_{\bm{k'},\bm{k},\bm{k'},\bm{k},s,s}^{v,c,c,v \vphantom{\dagger}} \right)
		- \left( g_{\bm{k},s,\bm{\xi_{L}}}^{v,c \vphantom{\dagger}} \right)^{\ast} n^{}_{\bm{\xi_{L}}}(t)
		\right)
	\nonumber \\
		\phantom{=}
		&{-} \left( \kappa_{\xi_{L}} + \Gamma \right) \psi^{}_{\bm{\xi_{L}},\bm{k},s}(t)
	\quad, \\
	\frac{\text{d}}{\text{d}t} f^{c}_{\bm{k},s}(t)
		=
		&{-} 2 \text{Re} \left[ \sum_{\bm{\xi_{L}}} g_{\bm{k},s,\bm{\xi_{L}}}^{v,c \vphantom{\dagger}} \psi^{}_{\bm{\xi_{L}},\bm{k},s}(t) \right]
		-\gamma_{nl} f^{c}_{\bm{k},s}(t) \left( 1 - f^{v}_{\bm{k},s}(t) \right)
		-\gamma_{rel} \left( f^{c}_{\bm{k},s}(t) - F_{\bm{k},s}^{D,c} \right)
		+P F_{\bm{k},s}^{p} \left( 1 - f^{c}_{\bm{k},s}(t) \right)
	\quad, \\
	\frac{\text{d}}{\text{d}t} f^{v}_{\bm{k},s}(t)
		=
		&\phantom{{-}} 2 \text{Re} \left[ \sum_{\bm{\xi_{L}}} g_{\bm{k},s,\bm{\xi_{L}}}^{v,c \vphantom{\dagger}} \psi^{}_{\bm{\xi_{L}},\bm{k},s}(t) \right]
		+\gamma_{nl} f^{c}_{\bm{k},s}(t) \left( 1 - f^{v}_{\bm{k},s}(t) \right)
		+\gamma_{rel} \left( F_{\bm{k},s}^{D,v} - f^{v}_{\bm{k},s}(t) \right)
		-P F_{\bm{k},s}^{p} \left( f^{v}_{\bm{k},s}(t) \right)
	\quad, \\
	\frac{\text{d}}{\text{d}t} n^{}_{\bm{\xi_{L}}}(t)
		=
		&\phantom{{-}} 2 \sum_{\bm{k'},s'} \text{Re} \left[ g_{\bm{k'},s',\bm{\xi_{L}}}^{v,c \vphantom{\dagger}} \psi^{}_{\bm{\xi_{L}},\bm{k'},s'}(t) \right]
		-2 \kappa_{\xi_{L}} n^{}_{\bm{\xi_{L}}}(t)
	\quad.
\end{align}
\end{widetext}

\section{Results and Discussion}

The QLEs have been solved on a hexagonal Monkhorst-Pack lattice with $60 \times 60 \times 1$ $\bm{k}$-points for a total of 21 pump rates
$P \in \left\{ 10^{-2.00},10^{-1.95},...,10^{-1.00} \right\}\,\text{ps}^{-1}$ until a stationary state has been reached at each pump rate.
The required calculations have been performed on the University of Bremen's QM3 cluster, utilising the LLNL's zvode \cite{zvode1988brown}
and a maximum of 20 physical Intel Xeon E5-2660 v3 2.60$\,$GHz processors (corresponding to 200 threads) for those calculations reaching
convergence in the nanosecond regime. The relevant simulation parameters can be found in TABLE 1.

\begin{figure}[h!]
\vspace{-0.25cm}
\includegraphics[width=0.50\textwidth]{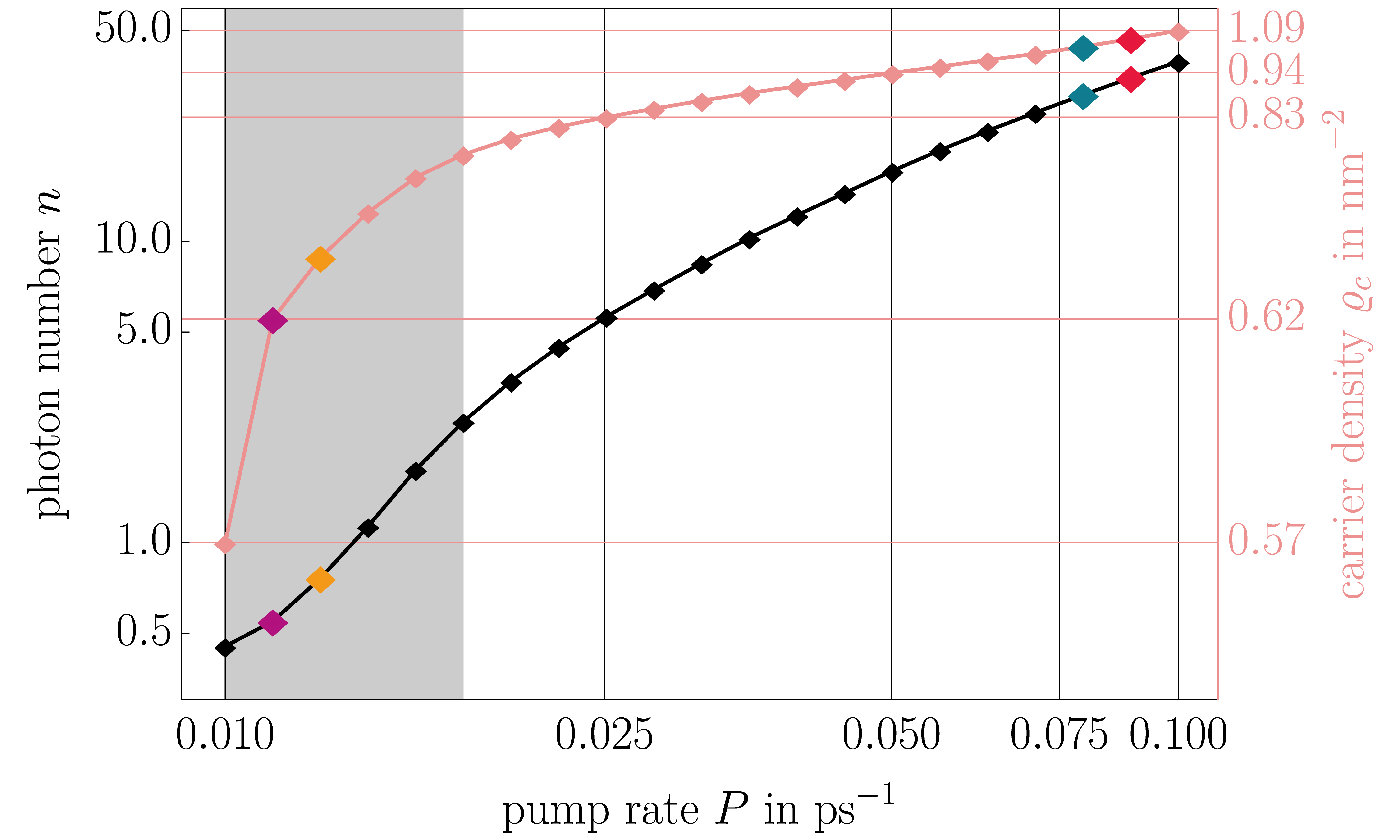}
\caption{Excitation-power dependant characteristics of the TMDC nanolaser, namely the photon number (input-output curve, black) and the overall carrier density (soft pink) in the optically relevant
conduction band. Exemplary pump rates are highlighted with purple ($P = 0.011\,\text{ps}^{-1}$), orange ($P = 0.013\,\text{ps}^{-1}$), blue ($P = 0.080\,\text{ps}^{-1}$), and red
($P = 0.089\,\text{ps}^{-1}$), respectively.}
\label{fig:B}
\vspace{-0.25cm}
\end{figure}

\begin{figure*}[ht!]
\vspace{0.00cm}
\includegraphics[width=1.00\textwidth]{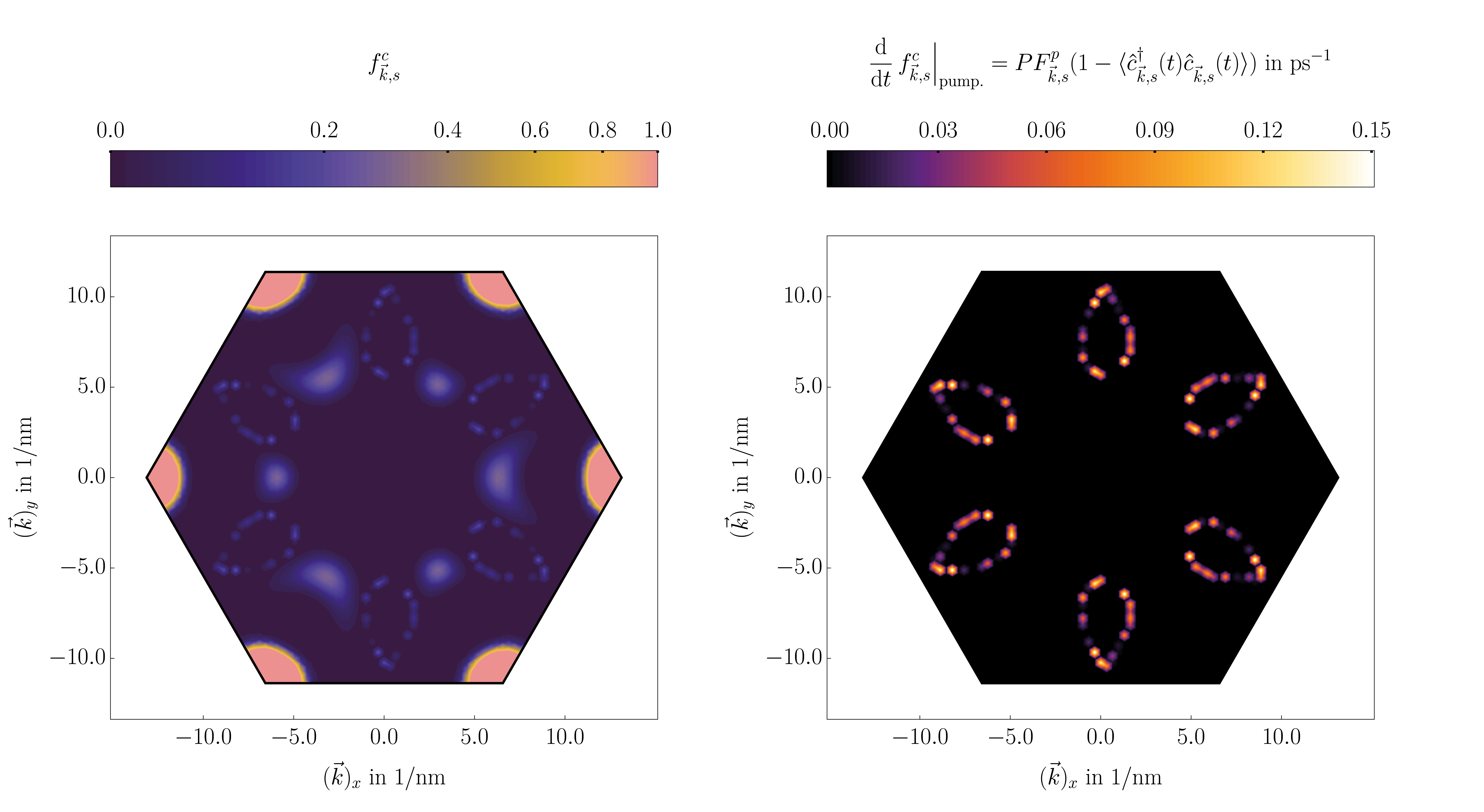}
\caption{Left: Stationary carrier population in the optically relevant conduction band at a pump rate of $P = 0.089\,\text{ps}^{-1}$ (red marker in other plots).
Right: Contributions from the pumping term to the build-up of carriers in the optically relevant conduction band. Here the spin-up case is shown.}
\label{fig:C}
\vspace{0.00cm}
\end{figure*}

The material-realistic microscopic modelling procedure employed here is highly cost-intensive from a numerical point of view, even on the doublet level.
Hence the second-order photon correlation function $g^{(2)}(\tau = 0)$ \cite{scully1997quantum,kira2011semiconductor,jagsch2018quantum,koulas2022quantum},
which can be calculated with the quadruplet level QLEs, has not been assessed. However, the theory presented here gives access to physical quantities with
both, high temporal and high momentum resolution, that allows the identification of strong indications for stimulated emission activity of the gain material.
In terms of small volume occupation and geometry, TMDC monolayer nanolasers share some properties with (Multiple) Quantum Well (MQW) nanolasers. However, an initial finding of this work is that
while MQW nanolasers exhibit a difficult to identify shallow threshold \cite{chow2014emission,kreinberg2020thresholdless,koulas2022quantum}, the input–output characteristics presented in FIG. 2
show a clearly visible s-shape in the double-logarithmic plot with the threshold being located between $P = 0.010\,\text{ps}^{-1}$ and $P = 0.018\,\text{ps}^{-1}$.
Between these pump rates, the photon number increases by one order of magnitude, coinciding well with an increase in the overall density of excited carriers in the optically relevant conduction
band. The time required to reach a stationary state at different pump rates illustrates the multi-scale nature of the dynamics: in the regime dominated by spontaneous emission (all pump rates up
to and including the orange marker), convergence times are in the order of magnitude of nanoseconds, while the stimulated emission regime is characterised by convergence times in the order of
magnitude of tens of picoseconds (compare FIG. 6, top). Notably, the timescale associated with convergence, i.e. the time required until a stationary state has been reached, undergoes a form of
phase transition by changing the order of magnitude once the photon number \smash{$n^{}_{\bm{\xi_{L}}}(t) \geq 1$}.
Furthermore, FIG. 2 illustrates that the excited carrier density does not exhibit clamping, hinting at a gain material heating effect at higher pump rates.
The distribution of carriers in the conduction band at these pump rates (FIG. 3, left) shows an accumulation of carriers in close proximity to and at the high-symmetry points $K$ and $K'$ at the
edges of the Brillouin zone, as well as $\Lambda$ and $\Lambda'$, which are local energy minima between the $\Gamma$-point and $K$ and $K'$, respectively.
This carrier accumulation is in conjunction with the energy landscape with one distinction: the three-fold symmetry also exhibited by the corresponding Fermi distributions at the
same densities is amended by addition of carrier populations arranged in a pattern resembling petals.

\renewcommand{\arraystretch}{1.5}
\begin{table}[h!]
\begin{ruledtabular}
\begin{tabular}{ll}
parameter & value \\ \hline
light-matter coupling strength 			& $g_{0} = 3.250\,(\text{nm\,pA\,ps}^2)^{-1}$ 			\\
resonator loss rate 					& $\kappa_{\bm{\xi_{L}}} = 0.725\,\text{ps}^{-1}$ 		\\
spontaneous emission rate 				& $\gamma_{nl} = 0.180\,\text{ps}^{-1}$ 				\\
relaxation rate 						& $\gamma_{rel} = 10.0\,\text{ps}^{-1}$					\\
dephasing rate 							& $\Gamma = 40.0\,\text{ps}^{-1}$						\\
lasing energy 							& $\hbar \omega_{\bm{\xi_{L}}} = 1374.9\,\text{meV}$	\\
lasing wavelength 						& $\lambda_{\bm{\xi_{L}}} = 901.7\,\text{nm}$			\\
simulation temperature					& $T = 300.0\,\text{K}$									\\
\end{tabular}
\end{ruledtabular}
\caption{Relevant numerical parameters utilised to obtain the results presented in FIG.s 2-6.}
\end{table}

\begin{figure*}[ht!]
\vspace{0.00cm}
\includegraphics[width=1.00\textwidth]{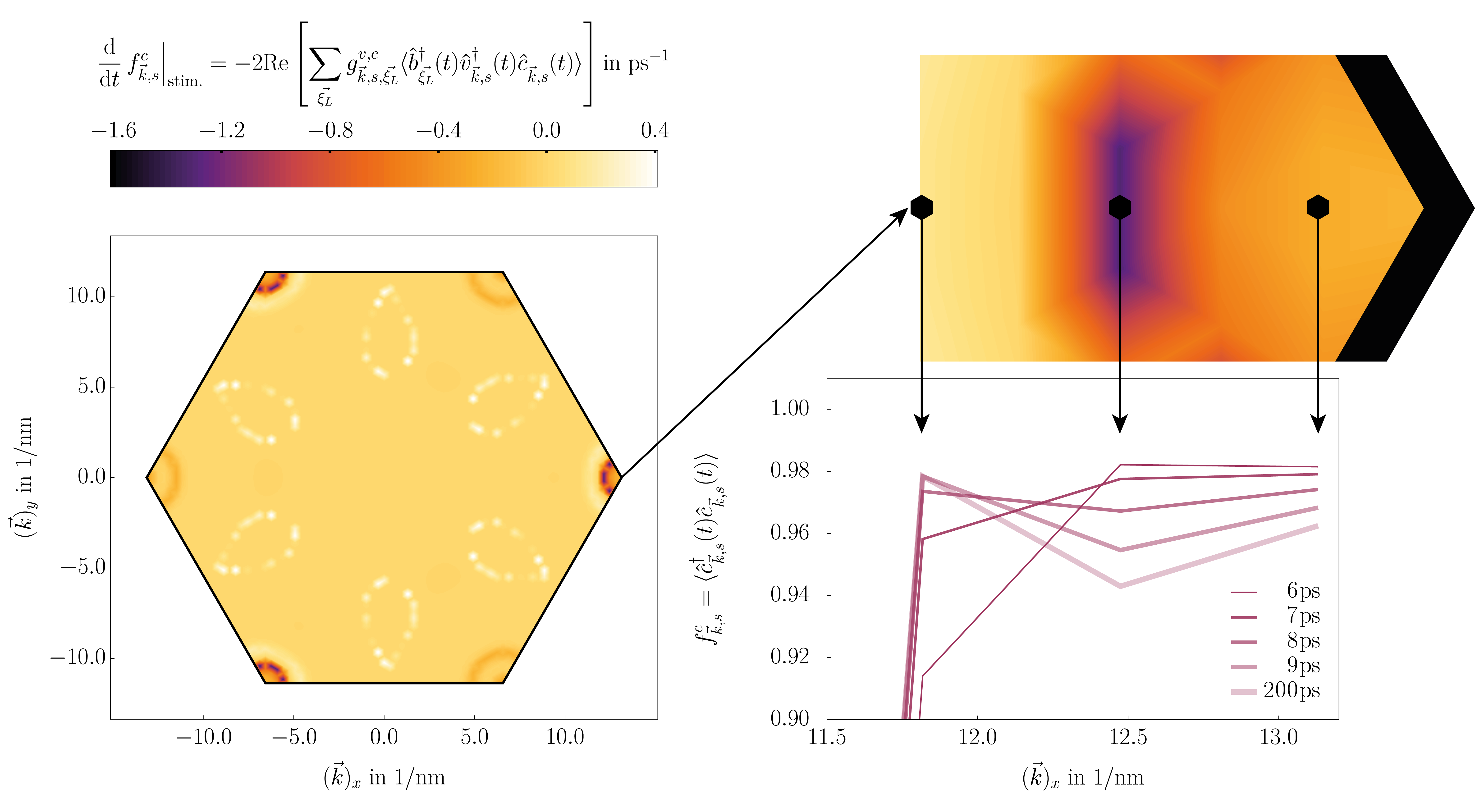}
\caption{Left: Contributions from the stimulated emission term to the reduction of carriers in the optically relevant conduction band at a pump rate of $P = 0.089\,\text{ps}^{-1}$
(red marker in other plots). Right, top: Zoom-in of the area around the $K$-point with three distinct neighbouring points of the employed Monkhorst-Pack-lattice at $(\bm{k})_{y} = 0$.
Right, bottom: Temporal evolution of the carrier population in this conduction band at these three lattice points for 6-9$\,$ps (build-up), and 200$\,$ps (stationary), respectively.
Here the spin-up case is shown.}
\label{fig:D}
\vspace{0.00cm}
\end{figure*}

\begin{figure}[h!]
\vspace{0.00cm}
\includegraphics[width=0.50\textwidth]{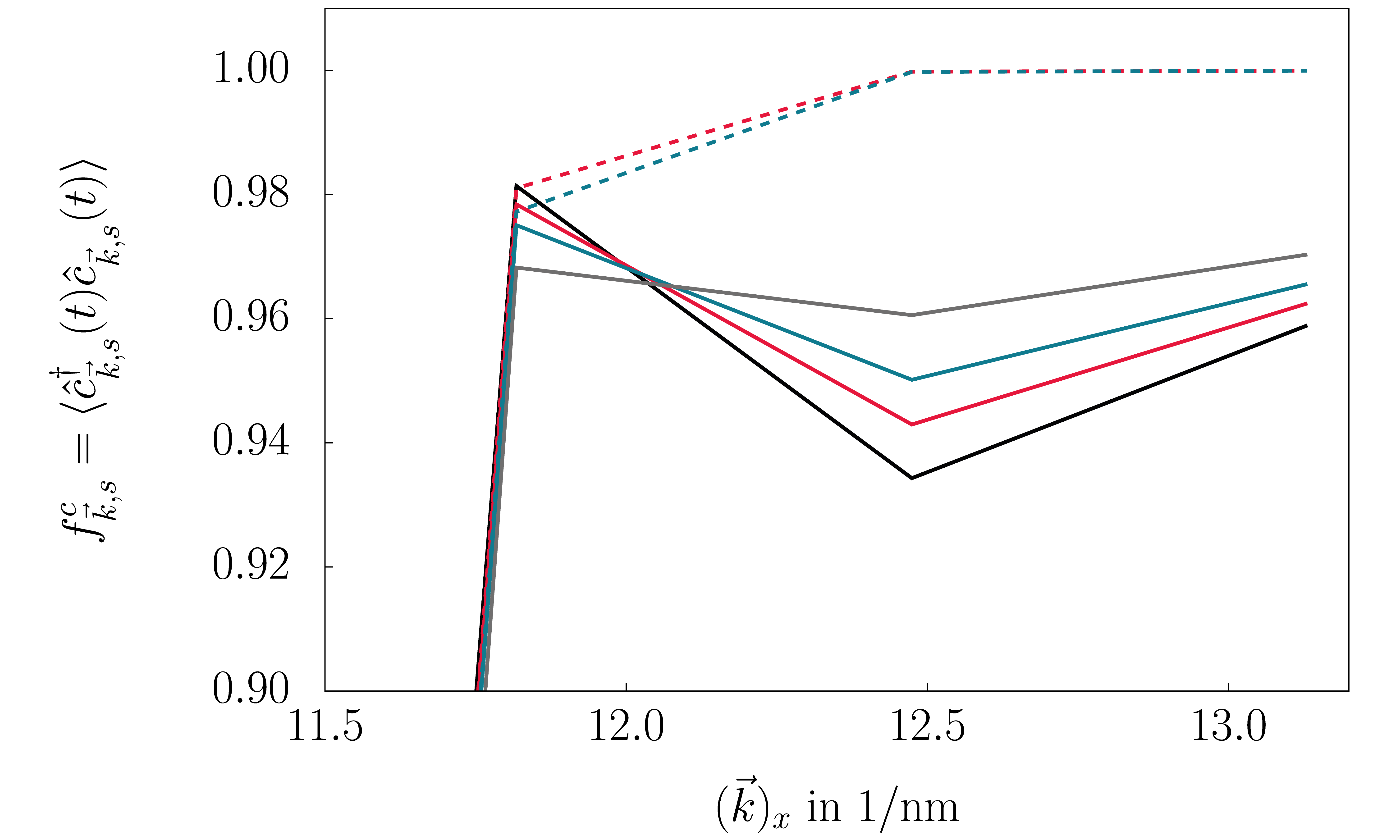}
\caption{Carrier population in the optically relevant conduction band at three distinct neighbouring lattice points in the area around the $K$-point for four different pump rates:
$P = 0.100\,\text{ps}^{-1}$ (black), $P = 0.089\,\text{ps}^{-1}$ (red), $P = 0.080\,\text{ps}^{-1}$ (blue), and $P = 0.063\,\text{ps}^{-1}$ (grey), respectively;
dashed lines indicate the Fermi-distributed populations for the densities of the corresponding colours. Here the spin-up case is shown.}
\label{fig:E}
\vspace{0.00cm}
\end{figure}

\noindent These petals originate from the pumping term occurring in the equation for the conduction band population (FIG. 3, right).
In the very same equation, the stimulated emission term shows a dominant negative rate around the $K$-point for the spin-up case ($K'$ for the spin-down case), highlighting that mainly
carriers located in close proximity to or at these high symmetry points are the ones contributing to the lasing activity of the device (FIG. 4, left and right, top).
This behaviour is further emphasised by the fact that during the build-up of carriers, hole burning can be observed at these very points (FIG. 4, right, bottom) with the stationary
population deviating 5.7$\,\%$ (FIG. 5, red) and 5.0$\,\%$ (FIG. 5, blue), respectively, from the Fermi-distributed population levels at the same total carrier density.
Furthermore, the populations at these $\bm{k}$-points show clamping in the regime of stimulated emission with the magnitude of hole burning being proportional to the pump rate $P$ (FIG. 5).
The comparison with Fermi distributions highlights that effects indicative of lasing can be identified when having access to the carrier dynamics impacted by many-body activity, such as energy
band renormalisations, via the photon-assisted polarisation \smash{$\psi^{}_{\bm{\xi_{L}},\bm{k},s}$}.
The final strong indicator for lasing activity beyond the input-output curve and hole burning is the occurrence of spectral clamping in the gain spectra (FIG. 6). The gain spectra have been
generated using the populations originating from the doublet level QLEs as input quantities to the Semiconductor Bloch Equations (SBEs) \cite{steinhoff2014influence}. While structurally related,
the SBEs, as opposed to the QLEs, describe the electromagnetic field semi-classically and not quantum mechanically. As a consequence, it is possible to use these equations for obtaining the
material's linear response based on its polarisation, which is driven by the electromagnetic probe field and furthermore influenced by carrier populations for non-zero densities.
In linear response theory, the absorption is proportional to the imaginary part of the susceptibility, which in turn can be derived from the material's polarisation
\cite{haug2004quantum,florian2018dielectric}. For population inversion, i.e. lasing, the absorption is negative and thus referred to as gain of the material.
Using the carrier populations in the stationary regime from the QLEs as input for the SBEs gives access to the  gain spectra of the TMDC monolayer nanolaser described here.
FIG. 6 (bottom) illustrates this for four exemplary pump rates, two of which being attributed to the regime of stimulated emission.

\begin{figure}[h!]
\vspace{0.00cm}
\includegraphics[width=0.50\textwidth]{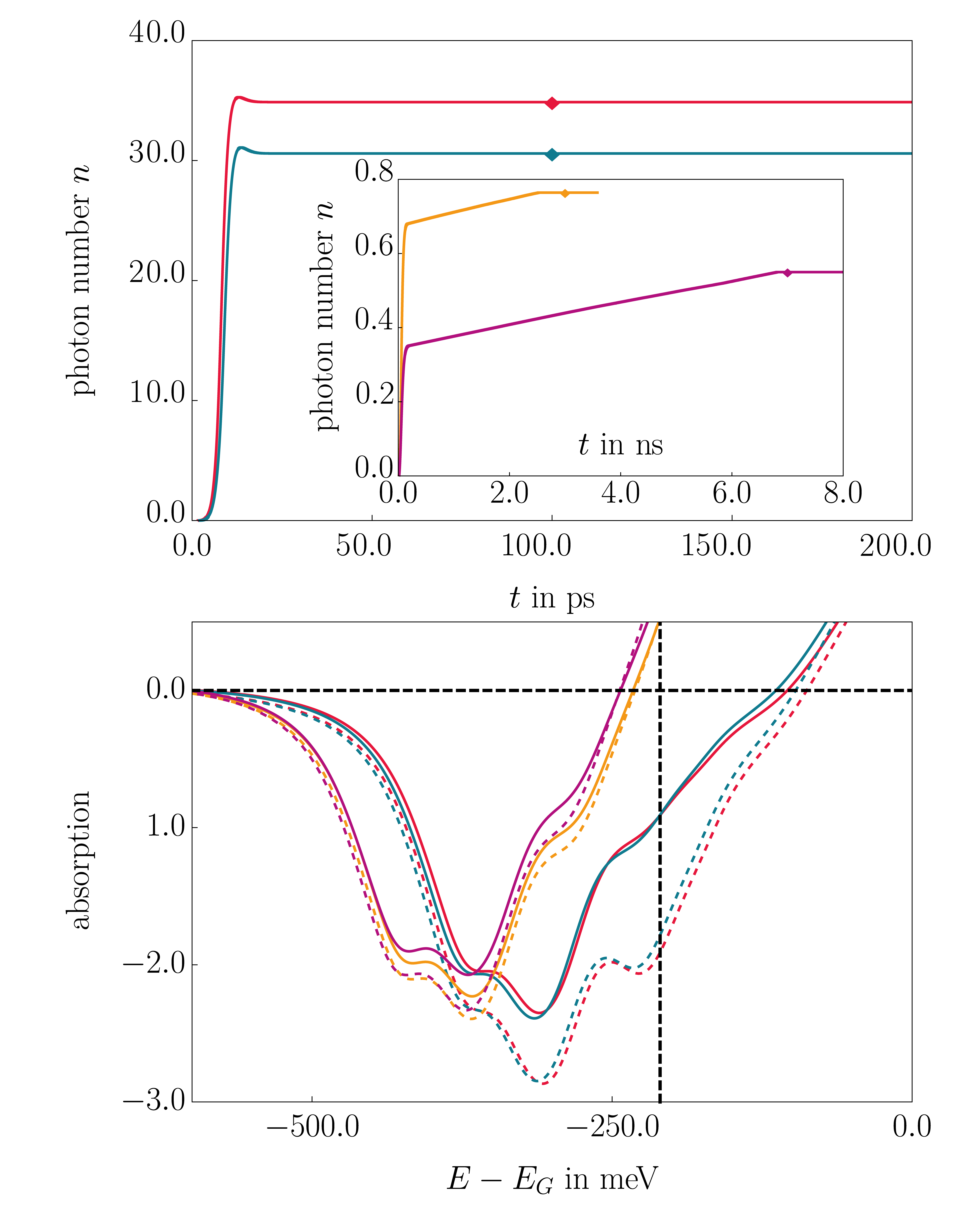}
\caption{Top: Temporal evolution of the photon number for four exemplary pump rates of the gain material. $P = 0.089\,\text{ps}^{-1}$ (red) and $P = 0.080\,\text{ps}^{-1}$ (blue) are in
the regime of stimulated emission while $P = 0.013\,\text{ps}^{-1}$ (orange) and $P = 0.011\,\text{ps}^{-1}$ (purple) are below the onset of this type of emission. \newline
Bottom: Absorption spectra generated using the Semiconductor Bloch Equations with identical parameters (if applicable) and matrix elements for the same four
exemplary pump rates using the populations originating from the Quantum Laser Equations extracted at the marker positions (solid lines) as well as Fermi-distributed populations for the
corresponding carrier densities (dashed lines). The dashed black lines indicate the transition to negative absorption, i.e. gain (horizontal), and the lasing mode energy of
$\hbar \omega_{\bm{\xi_{L}}} = 1374.9\,\text{meV}$ (vertical).}
\label{fig:F}
\vspace{0.00cm}
\end{figure}

\vspace{\baselineskip}

\noindent The associated absorption spectra cross at the lasing mode energy of $\hbar \omega_{\bm{\xi_{L}}} = 1374.9\,\text{meV}$ and demonstrate negative absorption, i.e. gain, above the
lasing threshold, hence emphasising that spectral clamping is exhibited by the presented TMDC nanolaser. The calculations reveal that even at a lasing frequency detuned -210$\,$meV against
the gap energy $E_G$, a spectral position not coinciding with the gain maximum, sufficient gain is present to sustain lasing activity. A direct comparison with Fermi distributions generated
for the same densities shows that the exact distribution of carriers is crucial for continuous stimulated emission at different pump rates. This is mostly due to the petal-shaped locations in
the Brillouin zone populated by the pumping term as well as the $K$- and $K'$-points including their close vicinity populated by scattering effects. If only Fermi distributions were generated
by the dynamics, no stable gain phase could be established as visualised by the offset of the red and blue dashed lines in FIG. 6 (bottom) at the lasing mode energy.

\section{Conclusion}

We have demonstrated the feasibility of solving the multi-scale problem that arises when microscopically modelling the many-body dynamics of a TMDC nanolaser device using a monolayer of MoS$_2$ as
gain material at room temperature. The characteristic s-shape of the input-output curve serves as an initial indicator of electron-hole-plasma-based lasing as described by our theory
utilising a fully quantised electromagnetic field, which is the method of choice for nanolasers with very low photon numbers, and microscopic Coulomb interactions. The lasing threshold is
accompanied by a shifting of the temporal dynamics with convergence times decreasing by one order of magnitude at the transition from the spontaneous to the stimulated emission regime.
Utilising the access to details regarding the reduction of carrier populations allows us to identify the $K$- and $K'$-points including their close vicinity as origins of carriers contributing
predominantly to lasing activity of the device. This is also emphasised by a pronounced hole burning process exhibited around these high-symmetry points. The consistently calculated absorption
spectra furthermore show that the device demonstrates spectral clamping at the lasing frequency and hence establishes a stable gain phase. It is for these reasons that MoS$_2$ can be recommended
as a viable gain material candidate for future opto-electronic applications like chips with integrated photonics.

\newpage


%

\end{document}